\documentclass[12pt,epsf]{article}
\newlength{\abstractwidth}
\usepackage[dvips]{graphicx}

\begin{document}

\flushbottom \thispagestyle{empty} \pagestyle{plain}

\renewcommand{\thefootnote}{\fnsymbol{footnote}}
\renewcommand{\thanks}[1]{\footnote{#1}} 
\newcommand{\starttext}{
\setcounter{footnote}{0}
\renewcommand{\thefootnote}{\arabic{footnote}}}
\renewcommand{\theequation}{\thesection.\arabic{equation}}
\newcommand{\be}{\begin{equation}}
\newcommand{\bea}{\begin{eqnarray}}
\newcommand{\eea}{\end{eqnarray}}
\newcommand{\beq}{\begin{equation}}
\newcommand{\ee}{\end{equation}}
\newcommand{\eeq}{\end{equation}}
\newcommand{\N}{{\cal N}}
\newcommand{\<}{\langle}
\newcommand{\ophi}{{\cal O}_\phi}
\newcommand{\oc}{{\cal O}_C}
\renewcommand{\a}{\alpha}
\renewcommand{\b}{\beta}
\newcommand{\m}{\mu}
\newcommand{\n}{\nu}
\newcommand{\tp}{\tilde p}
\newcommand{\mf}[1]{m_\phi^2(k_{#1})}
\newcommand{\mt}[1]{m_t^2(k_{#1})}
\newcommand{\half}{{1\over 2}}
\newcommand{\cabc}{\<C^{k_1}C^{k_2}C^{k_3}\>}
\newcommand{\D}[1]{\Delta_{#1}}
\newcommand{\tpm}[1]{\tilde p^{\mu_{#1}}}
\renewcommand{\>}{\rangle}
\def\ba{\begin{eqnarray}}
\def\ea{\end{eqnarray}}
\newcommand{\PSbox}[3]{\mbox{\rule{0in}{#3}\includegraphics{#1}\hspace{#2}}}

\def\N{{\cal N}}
\def\O{{\cal O}}
\def\Re{{\rm Re}}
\def\Im{{\rm Im}}
\def\tr{{\rm tr}}
\def\12{{1 \over 2}}
\def\32{{3 \over 2}}
\def\72{{7 \over 2}}
\def\92{{9 \over 2}}
\def\d{{d \over 2}}
\def\k{{k \over 2}}
\def\AdS{{\rm AdS}}
\def\dmu{\partial _\mu}
\def\dnu{\partial _\nu}
\def\dmup{\partial _{\mu '}}
\def\dnup{\partial _{\nu '}}
\def\a{\alpha'}
\def\nc{non--commutative}
\def\ny{non--commutativity}
\def\snc{space/time non--commutative}
\def\sny{space/time non--commutativity}
\def\ft{fuzzy torus}
\def\udag{U^{\dag}}
\def\vdag{V^{\dag}}
\def\xdag{X^{\dag}}
\def\ydag{Y^{\dag}}
\def\zdag{Z^{\dag}}
\def\ep{e^{i \theta}}
\def\em{e^{-i \theta}}
\def\epp{e^{2i \theta}}
\def\epn{e^{i(N-1) \theta}}
\def\q{&=&}
\def\x{{\cal{X}}}
\def\y{{\cal{Y}}}
\def\xd{{\cal{X}}^{\dag}}
\def\yd{{\cal{Y}}^{\dag}}
\def\ap{area preserving diffeomorphism}
\def\aps{area preserving diffeomorphisms}
\def\ep{\epsilon^{ij}}
\def\dz{$D0$-brane}
\def\dzs{$D0$-branes}
\def\dt{$D2$-brane}
\def\dts{$D2$-branes}
\def\ds{$D6$-brane}
\def\dss{$D6$-branes}
\def\de{D8-brane}
\def\des{D8-branes}
\def\cs{Chern Simons}
\def\qp{quasiparticle}
\def\const{c} 
\def\del{\nabla}


\renewcommand{\theequation}{\thesection.\arabic{equation}}
\begin{titlepage}
\bigskip
\rightline{SLAC-PUB-8657} \rightline{SU-ITP 00-25}
\rightline{hep-th/0108076}

\bigskip\bigskip\bigskip\bigskip

\centerline{\Large \bf {A Two Fluid Description of the Quantum Hall Soliton}}

\bigskip\bigskip
\bigskip\bigskip

 \centerline{\it
B. Freivogel, L. Susskind and N. Toumbas }
\medskip
\centerline{Department of Physics}
\centerline{Stanford
University} \centerline{Stanford, CA 94305-4060}
\bigskip\bigskip
\begin{abstract}

We show that the Quantum Hall Soliton constructed in
\cite{giantbob} is
stable under small perturbations.  We find that
creating quasiparticles
actually lowers the energy of the system, and discuss
whether this
indicates an instability on the time scales relevant
to the problem.
\medskip
\noindent
\end{abstract}

\end{titlepage}
\starttext \baselineskip=18pt \setcounter{footnote}{0}

\setcounter{equation}{0}
\section{Introduction and Review}
In \cite{giantbob} a configuration of branes and
strings -- the
Quantum Hall Soliton -- was discussed with low energy
dynamics similar
to those of condensed matter systems displaying the
fractional Quantum
Hall effect. The configuration consists of a spherical
$D2$-brane
wrapping $K$ flat $D6$-branes. For topological reasons
$K$ fundamental
strings must stretch from the $D6$-branes at the
center to the
spherical
$D2$-brane. The string ends on the $D2$-brane play the
role of the
electrons. The magnetic flux quanta are $N$
$D0$-branes dissolved in
the $D2$-brane. The filling factor is therefore given
by
\be
\nu = {K \over N}.
\ee
Several phenomena that occur in Quantum Hall systems
can be modelled
qualitatively in terms of the strings and branes
involved in the
configuration.

In \cite{giantbob}, it was shown how to describe the
background
magnetic field in terms of an incompressible fluid of
$D0$-branes
using Matrix Theory. In this paper, we follow
\cite{susskind} and show
how to model the electrons as a charged fluid moving
with the
$D0$-brane
fluid. The resulting two-fluid description allows us
to investigate
further aspects of the Quantum Hall Soliton dynamics
at low
energies. The effective action describing the two
interacting fluids
involves two gauge fields coupled together with a
scalar field
controlling the size of the soliton. In the two-fluid
picture,
quasiparticles may be thought of as vortices in the
electron fluid.

It was shown in \cite{giantbob} how the soliton can be
stabilized in
the near horizon geometry of the $D6$-branes. It was
found that there
is a characteristic energy scale associated with the
low energy
dynamics of the soliton. The stability of the
configuration with
respect to a variety of perturbations was discussed
and it was found
that in the large $N$ limit the soliton is stable
under such
perturbations. An issue that was left open is the
stability of the
soliton under non-spherically symmetric perturbations
of the
configuration. A preliminary investigation of the
stability of the
soliton with respect to non-spherically symmetric
perturbations was
carried out in \cite{bena}. The description of the
soliton dynamics in
terms of the two fluids allows us to investigate
thoroughly the
stability of the
configuration. It is found that the system is stable
under such
perturbations. The energy scale for such oscillations
is of the same
order as the characteristic energy scale identified in
\cite{giantbob}.

Finally, we note that the creation of quasiparticles
(vortices in the electron fluid with charge equal to
the filling fraction) actually lowers the energy of
the system. Naively, this seems like an instability.
However, in the small perturbations regime we are
working
in there is a static solution with any number of
quasiparticles,
and a conservation law prevents the creation or
annihilation of quasiparticles. There are many higher order
terms in the Lagrangian which we ignore because we
assume small fluctuations, so we can only
speculate about whether these terms will lead to an
instability on a time scale relevant to the problem. We discuss
this matter further in the stability section.

Please note that the construction discussed in this
paper is different from the more recent construction of Hellerman and
Susskind \cite{HS}. Roughly speaking, the electrons and the
$D0$-branes swap roles in the two constructions.

\setcounter{equation}{0}
\section{The Two Fluids}
In this section, we follow \cite{susskind} and
describe the Quantum
Hall Soliton dynamics in terms of two coupled fluids,
one describing
the $D0$-branes and the other describing the string
ends. The fluid
descriptions are valid for large $N$ and $K$ and at
distances bigger
than the microscopic length scales of the problem.
These are the
string length scale $l_s$ and the magnetic length. As
we saw in
\cite{giantbob}, the magnetic length is of order the
string scale. We
focus on a flat $D2$-brane substrate first and
generalize to the
spherical brane
configuration appropriate for the Quantum Hall Soliton
later.

\subsection{The $D0$-brane Fluid}
When a $D0$-brane enters a $D2$-brane, it dissolves
into magnetic
flux. The density of the $D0$-branes is equivalent to
a magnetic field
on the membrane while the particle currents result in
an electric
field. To see the precise connection, we recall that
in $2+1$
dimensions the field strength $F_{\mu\nu}$ is dual to
a
$3$-vector $J^{\mu}$
\be
J^{\mu}= {1 \over 2}\epsilon^{\mu\nu\rho}F_{\nu\rho}.
\ee
Thus,
\be
J^{0}=B={1 \over 2}\epsilon^{0ij}F_{ij}=F_{12} \ee is
the density of
the
$D0$-branes $\eta$ and
\be
J^{i}=BV^i=\epsilon^{i0j}F_{0j} = -\epsilon^{ij}E_j
\ee
are the
particle currents
$\eta V^i$. Here, $V^i$ may be thought of as the
velocity field of the
fluid
particles. The Bianchi identity for the field strength
\be
\epsilon^{\mu\nu\sigma}\partial_{\mu}F_{\nu\sigma}=0
\ee
becomes the continuity equation for the $D0$-brane
fluid
\be
\partial_{\mu}J^{\mu}=\partial_t\eta+\partial_i(\eta
V^i)=0.
\ee

In order to desribe the $D2$-brane dynamics, one
usually employs the
static gauge in which the worldvolume co-ordinates
$\xi^0,\,\
\xi^1,\,\ \xi^2$ are set equal to the embedding fields
$X^0,\,\
X^1,\,\ X^2$ along the directions parallel to the
brane. In this
gauge, the co-ordinate freedom of the
problem--worldvolume
diffeomorphisms--is completely fixed. The dynamics is
then described
by a $U(1)$ gauge field $A_\mu(X)$--or in terms of the
gauge invariant
quantities $B(X)$ and $E_i(X)$--in addition to the
embedding
fields along the
directions transverse to the brane. From the point of
view of the
fluid picture, the meaning of this gauge is clear: we
use co-ordinates
$X^0=t,\,\ X^1,\,\ X^2$ fixed in space and the
equations of motion of
the fluid are expressed in terms of the density and
the velocity
fields, $\eta(X)=B(X)$ and $V^i(X)=-\epsilon^{ij}E_j(X)/B(X)$. This
is what is
usually called the ``Eulerian description'' of the
fluid.

In fluid dynamics, there is another description of the
fluid, called
the ``Lagrangian description''. In this description,
we use
co-ordinates $\xi^0=t$ and $\xi^1=y^1,\,\ \xi^2=y^2$
co-moving with the
fluid. In this frame, the density of the fluid
$\eta$ is fixed
and the currents $\eta V^i$ are zero. From the point
of view of the
$D2$-brane theory, we may work in a frame such that
the magnetic field
$B$
is fixed to a constant value and the electric field
$E_i$ is
zero \cite{cornalba}. Equivalently, the field strength
$F_{\mu\nu}$ is
constant with
\be
F_{ij}=B_{ij}\ee
where $B_{ij}$ is equal to the constant matrix $B
\epsilon_{ij} $, and
\be
F_{0i}=0.\ee

Thus in this particular frame, the gauge field on the
membrane can be taken to be fixed and non-fluctuating.
For example, we
can work in the $A_0=0$ gauge and set
\be
A_i=-{B \over 2}\epsilon_{ij}y^j.
\ee
The dynamical
fields are the embedding fields $X^i(y,t)$ along the
directions
parallel to the brane and the embedding fields along
the directions
transverse to the brane.

The requirement that the field strength is constant
does not
completely fix the co-ordinate freedom of the problem.
One may find
co-ordinate transformations that leave the two form
$F_{ij}=B_{ij}$
constant and $F_{0i}$ zero. Infinitesimally, such
transformations
take the form
\be
y'^i=y^i+\Theta^{ij}\partial_j\lambda,
\ee
where
\be
\Theta^{ij}=(B^{-1})^{ij}=-B^{-1}\epsilon^{ij}
\ee
and $\lambda$ is time independent. Such
transformations are time
independent area preserving diffeomorphisms.

Under a co-ordinate transformation the embedding
fields $X^i$
transform as scalars,
\be
X'^i(y')=X^i(y).
\ee
Therefore, under infinitesimal area preserving
diffeomorphisms,
eq. (2.9),
\be
\delta X^i(y)=
\Theta^{jk}\partial_j\lambda\partial_kX^i=-i\{\lambda,
X^i\},
\ee
where the Poisson bracket between two quantities $A$
and $B$ is
defined by
\be
\{A, B\}=i\Theta^{ij}\partial_iA\partial_jB.
\ee

For studying small oscillations of the fluid, it is
appropriate to
introduce the displacement fields $\hat{A_i}(y,t)$
defined by
\be
X^i=y^i-\Theta^{ij}\hat{A_j}.
\ee
Then under an infinitesimal area preserving
diffeomorphism
\be
\delta\hat{A_i}=-\partial_i\lambda + i\{\lambda,
\hat{A_i}\}.
\ee
We conclude that the displacement fields are
non-commutative gauge
fields. Their transformations are the usual
non-commutative gauge
transformations \cite{seibergwitten} truncated to
first order in
$\Theta$. The full non-commutative gauge symmetry can
be recovered by
replacing the fluid description of the $D0$-branes by
their microscopic
Matrix
Theory description \cite{giantbob}\cite{seiberg}.

The ``Eulerian'' and ``Lagrangian'' descriptions of
the $D0$-brane
fluid are related to each other by a co-ordinate
transformation
\cite{cornalba}\cite{seiberg}. We review their
relation in the
appendix.
Here, we note that the ordinary gauge field $A_i$
of the ``Eulerian'' description is related to the
displacement field
$\hat{A_i}$ by the Seiberg-Witten transformation
between ordinary and
non-commutative gauge fields \cite{seibergwitten}.

{\bf Chern Simons Couplings.} Now let us see the
effect of turning on
a constant $D0$-brane magnetic field $H_2$ along the
directions of the
$D2$-brane. Let the total $H_2$-flux through the
$D2$-brane be $2\pi
\mu_6 K$ as in \cite{giantbob}. This flux is sourced
by the $D6$-branes. The total magnetic flux due to the
dissolved $D0$-branes is $2\pi N$.
Thus
\be {H \over B}= {\mu_6 K \over N}=\mu_6 \nu.
\ee

We can work in a gauge in which the
$RR$ gauge potential
$C_1$ is given by
\be
C_i=-{H\over 2}\epsilon_{ij}X^j.
\ee
Now the effect on the worldvolume theory of the
$D2$-brane is a
Chern-Simons type coupling given by
\be
{\mu_2 (2\pi \alpha') \over 2}\int d^3\xi
\epsilon^{\alpha \beta
\gamma}C_\mu(X)
{\partial X^{\mu} \over \partial
\xi^{\alpha}}F_{\nu\rho}(X){\partial
X^{\nu} \over \partial \xi^{\beta}}{\partial X^{\rho}
\over \partial
\xi^{\gamma}}.
\ee
Choosing to work with co-moving co-ordinates $t, \,\
y^1, \,\ y^2$ in
which the field strength is constant, the coupling
becomes\footnote{In our conventions,
$\epsilon^{012}=\epsilon^{12}=1$.}
\be
{\mu_2(2\pi \alpha') \over 2}\int dtd^2y
\epsilon^{0ij}C_k(X)
{\partial X^{k} \over \partial t}B_{ij}={\mu_2 (2\pi
\alpha') H B \over
2}\int dtd^2y \epsilon_{ij}X^i
{\partial_t X^{j}}.
\ee
This term has an intuitive explanation: a single
$D0$-brane (labeled by
an index $\alpha$) moving in
a constant magnetic field will have a term in its
Lagrangian
proportional to
\be
\sim {H \over
2}\epsilon_{ij}x_{\alpha}^i\partial_tx_{\alpha}^j.
\ee
Now in describing the many $D0$-brane system as a
fluid, $x_{\alpha}^i$
are replaced by the fields $X^i(y)$ and the sum over
all such
particles,
$\sum_{\alpha}$, by the integral $\int d^2y B$, where
$B$ is the
density of the particles.

In terms of the displacement fields, this term becomes
\be
-{\mu_2(2\pi \alpha')H \over 2} \int dtd^2y
\Theta^{ij}\hat{A_i}
{\partial_t \hat{A_j}} =  {\kappa \over 4\pi}\int dt
d^2y\epsilon^{ij}\hat{A_i}
{\partial_t \hat{A_j}},
\ee
up to a total time derivative. So it becomes an
ordinary Chern-Simons
coupling for the displacement fields. The level
$\kappa$ is given by
\be
\kappa = 2\pi \mu_2 (2\pi \alpha') H B^{-1}= 4\pi^2
\alpha' \mu_2 \mu_6
{K \over N} = {K \over N}=\nu.
\ee

{\bf Equation of constraint.} Since we are working in
the $A_0=0$
gauge, we must impose the $A_0$ equation of motion as
a
constraint. From
(2.18), the part of the Lagrangian containing $A_0$ is
given by
\be
{\mu_2(2\pi \alpha')H\over 2
}\epsilon^{ij}\epsilon_{kl}\partial_i X^k
\partial_j X^l A_0
\ee
and so the contribution to the equation of motion is
given by
\be
{\mu_2(2\pi \alpha')H \over 2}
\epsilon^{ij}\epsilon_{kl}\partial_i
X^k
\partial_j X^l + ... =0.
\ee
This term has the following origin.
In the static gauge, the effect of the $D6$-branes at
the center is to
couple to the $A$ field just like a uniform charge
density given by
\cite{giantbob}
\be
J^0={\mu_2(2\pi \alpha')H \over 2}.
\ee
Then, the corresponding density in the comoving frame
is given by
\be
J^0 \left({\partial X \over \partial y}\right)
\ee
where the last factor is the Jacobian of the
transformation from the
comoving frame to the ``Eulerian frame'' of the static
gauge. Expanding to linear order in the displacement
fields, we obtain
\be
J^0 + {\nu \over 2\pi}\epsilon^{ij}\partial_i\hat{A_j}
+ ... =J^0 +
{\nu \over
2\pi}\hat{F}_{12}+ ... =0.
\ee
Here
\be
\hat{F}_{ij}=\partial_i\hat{A_j}-\partial_j\hat{A_i}.
\ee
Including the non-linear terms in eq. (2.24) amounts
to replacing $\hat{F}$ with
\be
\hat{F}_{ij}=\partial_i\hat{A_j}-\partial_j\hat{A_i}-i\{\hat{A_i},
\hat{A_j}\},
\ee
i.e. with the non-commutative field strength.

The equations of motion, together with the constraint
equation, can be
derived by introducing a scalar potential $\hat{A_0}$
and varying the
following action \footnote{We have dropped total
derivatives.}
\be
S=\int dtd^2y\left[ -\hat{A_0}J^0 - {\nu \over
4\pi}\epsilon^{\mu\nu\rho}\hat{A_\mu}\partial_{\nu}\hat{A_\rho}\right].
\ee
The first term is a chemical potential term for
$\hat{A_0}$. It
acts like an induced positive background charge
density on the
membrane.  As we shall see in the next section, it can
be cancelled by
adding $K$ string ends on the membrane of opposite
charge:
\be
\int J^0 = K.
\ee
The second term is a Chern-Simons coupling. We have
kept terms to
quadratic order in the displacement fields only. Both
terms have been
derived in \cite{giantbob} using Matrix Theory with
the full
non-commutative gauge symmetry manifest. As we shall
see in the
next section, however, the fluid of string ends has
the effect of
cancelling both terms including the Chern-Simons
coupling.

What we have said above applies to the case of the
spherical Quantum
Hall soliton as well. In this case, the co-moving
co-ordinates $y^1,
\,\ y^2$
are two angles $\cal{\theta}$ and $\cal{\phi}$ and the
fixed density of
the
$D0$-brane fluid is given by
\be
B=B_{12}={N\over 2}  \sin{{\cal\theta}}.
\ee
The symmetry of the problem is the group of area
preserving
diffeomorphisms of the sphere. Under an infinitesimal
area preserving
diffeomorphism, the density $B$ transforms covariantly
\be
B \rightarrow B'={N\over 2}\sin{\cal{\theta'}}.
\ee
The effect of the $D6$-branes at the center is to
induce a chemical
potential term and a Chern-Simons coupling for the
displacement fields
as we have argued above. The "charge density" is now
given by
\be
J^0 = {K\over 4\pi}\sin{\cal{\theta}}.
\ee

{\bf Born-Infeld Dynamics.} The Quantum Hall soliton
lives in the near
horizon geometry of a stack of $K$ $D6$-branes. The
background metric
and dilaton fields are given by
\be
ds^2 = \sqrt{\rho \over l_s}(dt^2 -
dy^ady^a)-\sqrt{l_s \over \rho}(d\rho^2 +
\rho^2d\Omega_2^2),
\ee where $\rho$ is a radial coordinate, and
\be
g_s^2 e^{2\Phi} = {4 \over K^2} \left({\rho \over
l_s}\right)^{3
\over 2}. \ee
The soliton can be stabilized at a co-ordinate
distance given by
\cite{giantbob}
\be
\rho_* = {({\pi N})^{2 \over 3} \over 2}l_s \ee for
all $N$ and
$K$.
The values of the induced metric $g_{\mu\nu}$ on the
$D2$-brane and
dilaton field at $\rho_{*}$ are given by
\be
ds^2_{ind} = {(\pi N)^{1 \over 3} \over
\sqrt{2}}dt^2-{(\pi N)\over
2\sqrt{2}}\alpha'(d\theta^2 + \sin^2{\theta}d\phi^2)
\ee
and
\be
g_se^{\Phi}|_{\rho_*} = {2 \over K} \left({\rho_*
\over l_s}\right)^{3 \over 4} = 2^{1 \over
4}{\sqrt{\pi N} \over
K} \ee independent of $g_s$ at infinity.

The proper area of the stable soliton is given by
\be
A= 4\pi \sqrt{{\rho_*}^3l_s} =\sqrt{2}\pi^2 N l_s^2.
\ee Thus
there is a universal density of $D0$-branes which is
of order one
in string units for all $N$ and $K$. This means that
the fluid of
$D0$-branes is incompressible. The separation of the
$D0$-branes is
the magnetic length and this is of order one in string
units.

The action for the $D2$--brane in the background
geometry is given as
usual by
the Dirac Born Infeld (DBI) action. In terms of
co-moving
co-ordinates in which the gauge field on the brane is
non-fluctuating,
the Born-Infeld part of the action takes the following
form
\be
S_{D2}=-{1 \over 4\pi^2 g_s l_s^3}\int dtd^2y
e^{-\Phi}det^{1 \over 2}[h_{\mu\nu}+2\pi\alpha'
B_{\mu\nu}],\ee
with $B_{0i}=0$ and $B_{12}= N\sin{\cal{\theta}}/2$.
Here,
$h_{\mu\nu}$ is the induced metric on the brane and it
depends on the
dynamical fields $X^i(y, t)$.

We may use the DBI action to obtain an effective
action for small
oscillations of the soliton about the spherical
equilibrium
configuration at
$\rho_{*}$. To do so, we expand the action to
quadratic order in the
displacement fields $\hat{A_i}$ and the fluctuation of
the radial field
$\chi$ defined by
\be
\rho(X)=\rho_{*}+2\pi\alpha' \chi(X).
\ee
For later convenience, we take the radial field $\chi$ to be a
function of the space-fixed co-ordinates $X^i$. Naturally $\chi$
transforms as a scalar under co-ordinate transformations
\footnote{that is,
$\chi_y(y)=\chi(X)=\chi(y)-\partial_i\chi(y)\Theta^{ij}\hat{A_j}$.}.
We do not consider motions along the directions
parallel
to the $D6$-branes because there is translational symmetry in these
directions and so they are uninteresting. The effective
action governs the dynamics of the
soliton at distance scales comparable to the size of
the
sphere. Alternatively, the effective action describes
the dynamics of
the $D0$-brane fluid at macroscopic length scales,
distances bigger
than the magnetic length.

Expanding the DBI action and writing down only the
terms
quadratic in the fields, we obtain
\be
S_{gaugefield}=-{1 \over 2g_{YM}^2}\int
dtd^2y(detG)^{1 \over 2}
\left[G^{00}G^{ij}\partial_t\hat{A_i}\partial_t\hat{A_j}
+
{1 \over
2}G^{ik}G^{jl}\hat{F}_{ij}\hat{F}_{kl}\right],
\ee
\be
S_{scalar}={1 \over 2g_{YM}^2}\int dtd^2y(detG)^{1
\over
2}|g_{\rho\rho}|\left[G^{00}(\partial_t\chi)^2 +
G^{ij}\partial_i\chi
\partial_j\chi - {16\sqrt{2} \over 9\pi N
\alpha'}\chi^2\right]
\ee
and the interaction piece
\be
S_{int}={8\nu \over 9\pi}\int dtd^2y\chi
\hat{F}_{12}.
\ee
The indices are contracted with the effective ``open
string'' metric
$G_{\mu\nu}$. In terms of the ``closed string'' metric
at $\rho_{*}$,
eq. (2.38), this is given by
\be
G_{00}=g_{00}(\rho_{*}), \,\
G_{ij}=g_{ij}(\rho_{*})-2\pi\alpha'B_{ik}g^{kl}(\rho_{*})B_{lj}.
\ee
Therefore,
\be
ds_{open}^2 = {(\pi N)^{1 \over 3} \over
\sqrt{2}}dt^2-{9(\pi N)\over
2\sqrt{2}}\alpha'(d\theta^2 + \sin^2{\theta}d\phi^2).
\ee
The gauge coupling constant $g_{YM}$ is given in terms
of the
effective ``open string'' coupling constant as follows
\be
g_{YM}^2=G_s(\rho_{*})l_s^{-1},
\ee
where
\be
G_s(\rho_{*})=g_s(\rho_{*})\left({detG_{\mu\nu} \over
det(g_{\mu\nu}+2\pi\alpha'B_{\mu\nu})}\right)^{1 \over
2}.
\ee
Therefore,
\be
g_{YM}^2=3 {2^{1 \over 4}\sqrt{\pi N} \over
K}l_s^{-1}=3 {2^{1 \over 4}\sqrt{\pi} \over
\nu \sqrt{N}}l_s^{-1}.
\ee

The contribution to the constraint equation, eq.
(2.27), from the
Born-Infeld term is given by
\be
{1 \over g_{YM}^2}\partial_i\left({(detG)}^{1 \over
2}
G^{00}G^{ij}\partial_t\hat{A_j}\right)+J^0 + {\nu
\over
2\pi}\hat{F}_{12}+...=0.
\ee
The new term in the equation is simply the analogue of
minus the
divergence
of the electric field in ordinary electrodynamics.

Finally, as in \cite{gubser}, let us do a conformal
transformation and
rescale the fields so that the metric takes the form
\be
d\tilde{s}^2 = dt^2-{9(\pi N)^{2/3}\over
2}\alpha'(d\theta^2 + \sin^2{\theta}d\phi^2)
\ee
and the fields appear with canonically normalized
kinetic terms in the
action. To this end, we set
\be
\tilde{A_i}^2 = { 1 \over g_{YM}^2
\sqrt{G_{00}}}\hat{A_i}^2, \,\
\tilde{\chi}^2={|g_{\rho\rho}|\sqrt{G_{00}}\over
g_{YM}^2}\chi^2={ 1
\over g_{YM}^2 \sqrt{G_{00}}}\chi^2
\ee
and express the action in terms of the rescaled
fields. We obtain the
following action
\be
S= \int dtd^2y(det\tilde{G})^{1 \over 2}\left(-{1
\over
4}\tilde{F}_{\mu\nu}^2 + {1 \over
2}(\partial_{\mu}\tilde{\chi})^2 -
{8 \over 9(\pi N)^{2/3}\alpha'}\tilde{\chi}^2\right)+
 {8 \over 3 (\pi
N)^{1/3}l_s}\tilde{\chi}\tilde{F}_{12}.
\ee

In these units, the size of the sphere is of
order
$N^{1/3}l_s$. Accordingly, the magnetic length is of
order
$N^{-1/6}l_s$. The
scalar field is massive with mass given by
\be
m_{\tilde{\chi}}= {4 \over 3(\pi N)^{1/3}l_s}.
\ee
So its Compton wavelength is comparable to the size of
the
sphere. Since the scalar field is massive, we conclude
as in
\cite{giantbob} that the soliton is stable under small
spherically
symmetric oscillations. The energy scale of these
oscillations is set
by the mass of the scalar field. As we shall see in
the next section,
the
Chern-Simons term for the displacement fields is
cancelled by the
collective motions of the electron fluid and we do not
include it in
the action. We also note that the supersymmetry
breaking scale for the
$D2$-brane theory is set by the size of the sphere
\cite{gubser}.

The rescaled fields couple to ordinary charges with
coupling constant
\be
g_{YM}G_{00}^{1/4}.
\ee
Therefore, one expects the strength of interactions
between string
ends to be of order
\be
\sim {1 \over \nu N^{1/3}}
\ee
in string units.

Focusing on a small patch of the sphere bigger than
the magnetic
length so that we can approximate it as flat, the
gauge kinetic term
is of the standard form
\be
-{1 \over 4}\int d^3x \tilde{F}^2
\ee
and so the speed of sound waves of the $D0$-brane
fluid is one
\be
c=1.
\ee
Thus the $D0$-brane fluid is stiff and difficult to
compress.

The interaction term between the scalar and the gauge
fields is also
noteworthy. Consider a non-spherically symmetric
deformation of the
soliton as in \cite{bena}. Then the $D0$-branes would
tend to
accumulate in the regions farther away from the
$D6$-branes where the
repulsion is less strong.  In other words, where
$\chi$ is positive ($D2$-brane farther from the
$D6$-branes), the energy is lowered by $F_{12}$
(excess
density of $D0$-branes) being positive, so that the
term
in the Hamiltonian will be $-\chi F_{12}$ and thus
$+\chi F_{12}$ in the Lagrangian.

\subsection{The Electron Fluid}
As we saw in \cite{giantbob}, $K$ fundamental strings
must stretch
from the $D6$-branes to the $D2$-brane due to the
Hanany-Witten effect
\cite{HW}. The strings will tend to distribute
themselves homogeneously so as to cancel the
background charge density
induced by the $D6$-branes on the membrane.
In what follows, we consider
a model in which
the strings remain in their ground state apart from
the motion of their
ends on
the $D6$- and $D2$-branes. As was argued in
\cite{giantbob}, the
strong gauge dynamics on the $D6$-brane side will tend
to bind the $K$
string ends into a ``baryon''. The properties of the
``baryon''
wavefunction under interchange of two strings would
then determine the
effective statistics of the string ends on the
$D2$-brane side.  We ignore the intrinsic statistics
of the string ends here.

When $K$ is large, and at distances larger than the
magnetic length,
we can treat the string ends on the $D2$-brane as a
fluid of
non-relativistic charged particles. We will take the
effective mass of the particles to be of order the
mass of a radially
stretched
string at rest
\be
m_{string}= {\rho_{*} \over 2\pi l_s^2}= {({\pi N})^{2
\over
3} \over
4\pi}l_s^{-1}.
\ee
More precisely, the effective string mass
above is computed by
assuming that the string
remains straight as its ends move around on the
branes. As discussed in
\cite{giantbob}, the energy scale of long string
oscillations is of the same order as the typical energy scale
controlling the dynamics of the soliton, much lower than the mass of a
stretched string at rest, and we do not expect them to
modify the effective mass of the electrons.
In the large $N$ limit the non-relativistic
approximation a good one.

{\bf Non-commutative charged particles.}
Consider a $D0$-brane fluid element fixed at the
origin in its rest
frame (the
$y$-frame) and denote the relative position of the
string end with
respect to it by ${y_s}^i(t)$. Then, the position of
the string end in
space is given by
\be
{x_s}^i(t)={y_s}^i(t) -\Theta^{ij}\hat{A_j}(y_s,t).
\ee
Similarly, the particle velocity is given by
\be
{d \over dt}{x_s}^i=\dot{{x_s}}^i=\dot{{y_s}}^i
-\Theta^{ij}\dot{\hat{A_j}}=\dot{{y_s}}^i-\Theta^{ij}\partial_t\hat{A_j}
-\dot{{y_s}}^k\Theta^{ij}\partial_k\hat{A_j}.
\ee

In the frame comoving with the $D0$-brane fluid, the
particle
interacts with a fixed background magnetic field and
so
\be
L_{gauge inter}= {eB\over
2}\epsilon_{ij}{y_s}^i\dot{y_s}^j.
\ee
We have chosen the background potential to be given by
$-B\epsilon_{ij}y^j/2$.
Using equations (2.61) and (2.62), we can write this
in terms of the
space fixed variables, $x_s$ and $\dot{x_s}$. Keeping
terms to
quadratic order in the fluctuations, the Lagrangian
(2.63) becomes
\be
L_{gauge inter}={eB\over
2}\epsilon_{ij}{x_s}^i\dot{x_s}^j-e\dot{{x_s}}^i\hat{A_{i}}(x_s)-{e
\over
2}\Theta^{ij}\hat{A_i}(x_s)\partial_t{\hat{A_j}}(x_s).
\ee
We see that the particle couples to the displacement
field as it
couples to the ordinary commutative gauge field but it
also feels a potential.

Under the infinitesimal area preserving
diffeomorphisms, eq. (2.9), the Lagrangian (2.63)
changes by a total
time derivative
\be
\delta L = - {e \over 2}{d \over dt}(\partial_i
\lambda y_s^i).
\ee
The action is invariant under such
co-ordinate transformations and so under the
non-commutative
gauge symmetry of the problem.

The above result has also an intuitive explanation.
Suppose we can ignore the kinetic term. Then the full
Lagrangian is given by (2.63) and
the equations of motion imply that
\be
{y_s}^i = constant.
\ee
That is, the particle is fixed with respect to any
$D0$-brane fluid
element and simply follows the fluid. This can also be
understood as
follows. Let $v^i$ denote the velocity of the particle
in fixed
space. Since the particle is
massless, the Lorentz force on the particle must be
set to zero
\be
E_i(x) + B(x) \epsilon_{ij} v^j = 0.
\ee
It follows that $v^i$ is equal to $-\epsilon^{ij}
E_j(x)/B(x)$
which according to (2.3) is simply the
velocity field of the $D0$-brane fluid evaluated at
the position of
the particle. Thus, the particle follows the fluid.

The rest of the Lagrangian involves the
non-relativistic kinetic
energy term
\be
L_{KE}= -{1\over 2}mg^{00}g_{ij}\dot{x_s}^i\dot{x_s}^j
\ee
and the interaction with the scalar field $\chi$
\be
L_{scalar inter}= -\chi(x_s).
\ee
The last term arises since the rest mass of a string
is proportional
to its length.

{\bf Fluid description.} We describe the many string
ends as a
fluid. To this end, we use co-ordinates $t$ and $z^1,
\,\ z^2$
comoving with the electron fluid\footnote{$z^1$ and
$z^2$ are two
angles $\theta$ and $\phi$}. In this frame, the
electron number
density is fixed and given by
\be
\eta_0={K\over 4\pi}\sin{\theta}.
\ee
To pass to the fluid description, we replace
${x_s}^i_{\alpha}$ by the
fields $X_s^i(z,t)$ and the sum $\sum_{\alpha}$ by the
integral $\int
d^2z\eta_0$.
We obtain the following effective action
\be
S_{strings}= \int dtd^2z{-\eta_0 m \over
2}g^{00}g_{ij}\partial_tX_s^i\partial_tX_s^j +
{e\eta_0B\over
2}\epsilon_{ij}{X_s}^i\partial_t{X_s}^j+ V,
\ee
where
\be
V=-e\eta_0\partial_t{X_s}^i\hat{A_{i}}+{e\nu \over
4\pi}\epsilon^{ij}\hat{A_i}\partial_t{\hat{A_j}}+\eta_0\chi\partial_i(X_s^i-z^i)
+h.\, o.
\ee
In the potential, we have kept terms up to quadratic
order in the
displacement field $\hat{A}$ and spatial and temporal
derivatives of
the field $X_s$. Since we are expanding about an
extremum of the
potential, any linear terms will cancel once we add
the two fluid
actions together. Thus we ignore them. As we already
mentioned in the
previous section, the string-end fluid Lagrangian
contains a
Chern-Simons coupling for the the displacements fields
$\hat{A}$ that
cancels the Chern-Simons coupling, eq. (2.21), if $e$
is negative. So we
can omit it.

The fluid action we have obtained has also an exact
gauge symmetry
which consists of time independent area preserving
diffeomorphisms
\be
z'^i=z^i + {\eta_0 \over
2\pi}\epsilon^{ij}\partial_j\lambda.
\ee
Under such transformations the fields $X_s^i$
transform as
scalars. The fields describing the $D0$-brane fluid
appear as
functionals of $X_s$ in this action and so they
transform as scalars.
Under such transformations the electron number density
transforms
covariantly.

For small oscillations of the charged fluid, it is
convenient to
introduce the displacement fields defined by
\be
X_s^i = z^i - {1 \over 2\pi \eta_0}\epsilon^{ij}D_j.
\ee
In analogy with the displacement fields describing
small oscillations
of the $D0$-brane fluid $\hat{A}$, the displacement
fields $D$ are
also non-commutative gauge fields. At the linearized
level, they
transform as ordinary gauge fields
\be
D_i \rightarrow D_i + \partial_i \lambda.
\ee

In terms of the displacement fields, the string fluid
action becomes
\be
S_{strings}= \int dtd^2z{- 1 \over
2g_e^2}(detG_e)^{1 \over
2}G_e^{00}G_e^{ij}\partial_tD_i\partial_t D_j
- {e\over
4\pi \nu}\epsilon^{ij}D_i\partial_tD_j -{e \over
2\pi}\epsilon^{ij}\hat{A_i}\partial_t D_j- {1\over
2\pi}\chi\epsilon^{ij}\partial_i D_j.
\ee
Notice the similarity with the $D0$-brane fluid
action. The first term
is a conventional Maxwell term. The effective metric
appearing in the
Maxwell term is given by
\be
G_e^{00}=g^{00}, \,\ G_e^{ij}=-{
1 \over (2\pi\eta_0)^2
(2\pi\alpha')^2}\epsilon^{ik}g_{kl}\epsilon^{lj}\sim
{1 \over \nu^2 N
\alpha'}
\ee
and the coupling constant by
\be
g_e ^2 = {1 \over (2\pi\alpha'^2) (2\pi\eta_0) m}
(detG)^{1 \over 2}
\sim {\nu \over \sqrt{N}l_s}.
\ee
The formulae are analogous to the formulae for the
effective open
string metric and coupling constant in the zero slope
limit
\cite{seibergwitten}! Here, the non-commutativity
parameter is set by
the number density of the electrons
\be
\Theta_e={1 \over 2\pi\eta_0}
\ee
and the Maxwell coupling constant is inversely
proportional to the
mass of the strings as in the case of the $D0$-brane
fluid, where the
coupling constant is inversely proportional to the
$D0$-brane mass.
Unlike the Maxwell term describing the $D0$-brane
fluid however, the
magnetic part of the Maxwell term is absent. This
means that unlike
the $D0$-brane fluid case, it does not cost much
energy to compress the
electron fluid.  Physically, what is going on is that
the electrons only interact with each other via the
electromagnetic field since we are ignoring their
intrinsic statistics, so the density of electrons only
comes into the action via the coupling to the $A$
field.

The second term is a Chern-Simons term for the gauge
field $D$. As in
\cite{susskind}, it arises in the fluid description of
charged
particles in a magnetic field. The presence of the
Chern-Simons term
means that the gauge field $D$ is massive with mass
given by
\be
m_D \sim {g_e^2 \over \sqrt{G_{00}}4\pi \nu} \sim
{\eta_0
\over \sqrt{g_{11}} m \nu \alpha' }
\ee
in the units we are working with. With the mass of the
string given by
eq. (2.60), we find that
\be
m_D \sim N^{-{1 \over 3}}l_s^{-1}.
\ee
In fact, this mass is really the cyclotron frequency
setting the
energy scale of higher Landau levels. As in
\cite{giantbob}, we see a
single energy scale describing the low energy dynamics
of the Quantum
Hall soliton.

The last two terms involve the interactions of the
electron fluid
gauge field with the $D0$-brane fluid gauge field and
the scalar
field. The interaction between the two gauge fields is
a Chern-Simons
coupling consistent with the two gauge symmetries of
the problem. The
origin of the interaction with the scalar field is
also easy to
understand. Consider a non-spherically symmetric
perturbation of the
soliton. Then the strings would tend to concentrate in
the region
closer to the $D6$-branes since their mass is
smaller there. The sign of
this coupling is opposite to that between the scalar
field and
$\hat{A}$. We will comment more on the interaction
terms in the next
two sections.

The contribution of the electron fluid to the
constraint equation,
eq. (2.51), can easily be obtained. The electron
density in the space fixed frame is given by
\be
\eta = \eta_0 \left({\partial z \over \partial
X}\right)= \eta_0 + {1
\over 2\pi}\epsilon^{ij}\partial_iD_j.
\ee
The density in the frame comoving with the $D0$-brane
fluid is
\be
\eta\left({\partial X \over \partial y}\right)= \eta_0
+ {1
\over 2\pi}\epsilon^{ij}\partial_iD_j + {\nu
\over 2\pi}\epsilon^{ij}\partial_i\hat{A_j}.
\ee
So the contribution to the constraint equation is
simply
\be
e\eta_0 + {e
\over 2\pi}\epsilon^{ij}\partial_iD_j + {e\nu
\over 2\pi}\epsilon^{ij}\partial_i\hat{A_j}.
\ee
With $e=-1$, the full constraint equation becomes
\be
{1 \over g_{YM}^2}\partial_i\left({(detG)}^{1 \over
2}
G^{00}G^{ij}\partial_t\hat{A_j}\right) + {e \over
2\pi}\epsilon^{ij}\partial_iD_j=0.
\ee
We see that both the background charge density $J_0$
and the
contribution from the Chern-Simons term, eq. (2.27),
cancel.

{\bf Conserved Quantities.} Under the transformation
eq. (2.75) the
Lagrangian of (2.76) changes by a total time
derivative
\be
{e \over 4\pi
\nu}\partial_t(\epsilon^{ij}\partial_i\lambda D_j).
\ee
Therefore, a conserved quantity exists and is given by
\be
\partial_i(\Pi^i_D - {e \over 4\pi \nu}\epsilon^{ij}
D_j),
\ee
where $\Pi^i_D$ is the momentum conjugate to $D_i$.
The conserved
quantity is therefore given by
\be
-{1 \over g_e^2}\partial_i\left({(detG_e)}^{1 \over 2}
G_e^{00}G_e^{ij}\partial_tD_j\right) - {e \over
2\pi \nu}\epsilon^{ij}\partial_iD_j +{e \over
2\pi}\epsilon^{ij}\partial_i\hat{A_j}.
\ee
In the absence of vortices we may consistently impose
\be
-{1 \over g_e^2}\partial_i\left({(detG_e)}^{1 \over 2}
G_e^{00}G_e^{ij}\partial_tD_j\right) - {e \over
2\pi \nu}\epsilon^{ij}\partial_iD_j +{e \over
2\pi}\epsilon^{ij}\partial_i\hat{A_j}=0
\ee
as a second constraint equation. Thus,
changes in the
$D0$-brane fluid density source an electric field for
the gauge field
$D$. In this way, $D0$-branes may be thought of as
charges under the
electron fluid gauge field! Adding a single $D0$-brane
corresponds to
adding a unit of magnetic flux. It follows then from
the above equation
that it
creates a fluctuation in the electron density of total
charge
$\nu$. As in \cite{giantbob}, additional $D0$-branes
can be thought of
as the Laughlin quasiparticles \cite{Laughlin}.

\setcounter{equation}{0}
\section{Stability Analysis}
In this section, we use the two-fluid action we have
obtained to show
that the Quantum Hall soliton is stable under a
variety of
perturbations. There are reasons to think that the
system might be
unstable. The velocity independent part of the
potential is given by
\be
V \sim \tilde{\chi}\epsilon^{ij}\partial_i
(D_j-b\tilde{A}_j),
\ee
with $b$ of order one, and it is not positive
definite. The physical
origin of the potential instability can be described
as follows
\cite{bena}. Consider a configuration in which the
$D6$-branes are
displaced away from the center of the spherical
membrane. The \dzs \  are
repelled by the \dss, while the electrons are
attracted to the \dss\
by the stretched strings. It is conceivable then that
with the
electrons moving closer and the \dzs\ farther away
from the \dss\ the
energy of the configuration gets lowered. This
argument on the other
hand
assumes that the electrons and the $D0$-branes move
independently of
each other, neglecting the velocity dependent forces
in the
potential. It is well known that charged particles
moving in a
magnetic field tend to follow magnetic flux lines, and
so there is
also a tendency for the electrons and the \dzs\  to
stick
together. The velocity dependent terms in the
potential are crucial
for the stability of the soliton.

In addition, the system might be unstable to
forming small ripples, with the
\dzs\  becoming concentrated on the peaks of the
ripples where they are farther from the \dss\ and the
electrons
becoming concentrated in the troughs where the strings
attached to them can be shorter.

Let us illustrate how the potential instability gets
removed with
a simple example. Consider the case of a charged
particle
in an upside-down harmonic oscillator potential moving
in the presence
of a uniform magnetic field along the z direction. The
Lagrangian is
given by
\be
L = \12 m \dot{\vec{X}}^2 + \12 k \vec{X}^2 +\12
eB \epsilon_{ij} X^i \dot X^j.
\ee
We will see that in spite of the tachyonic potential,
the system is stable for some values of the
parameters.
In terms of $Z=X^1+iX^2$, and ignoring the $X^3$
direction which is uninteresting, the Lagrangian
becomes
\be
L = \12 m \dot Z \dot Z^\star + \12 k Z Z^\star + {i
\over 4}eB (  \dot Z Z^\star -  Z \dot Z^\star).
\ee
The equation of motion is
\be
m \ddot Z + kZ + ieB \dot Z = 0
\ee
This is the equation of motion of a damped harmonic
oscillator.  Looking at the characteristic frequencies
shows that if
\be
(eB)^2 > 4km,
\ee
then the system oscillates;
otherwise it exhibits exponential growth or decay.
Physically this means
that if the $B$ field is large enough, then the
particle
has closed orbits; otherwise it rolls down the
potential hill. Therefore, for a certain range of
parameters the
extremum of the oscillator potential is stable. The
stability
condition (3.5) can also be written as follows
\be
\omega_{cycl} > \omega_{oscil},
\ee
where the cyclotron frequency is given by $|eB|/m$ and
the
characteristic harmonic oscillator frequency by
$(k/m)^{1/2}$. We
conclude that with all other quantities fixed, the
oscillator is
stable if the mass $m$ is not too large.

This example demonstrates that the stability of the
Quantum Hall
Soliton
depends on the relative size of the velocity dependent
terms to the
velocity independent terms in the potential. The
strength of the
velocity dependent terms is set by the cyclotron
frequency or the mass
of the gauge field $D$
\be
\omega_{cycl} \sim N^{-1/3}l_s.
\ee
The strength of velocity independent terms is set by
the product of the
gauge coupling constants of each fluid
\be
\sqrt{g_eg_{YM}}G_{00}^{1/4}\sim N^{-1/3}l_s.
\ee
Since these are comparable for all $N$ and $K$, a
careful analysis of
the stability of the configuration is needed.

We first worry about perturbations on the scale of the
whole sphere. Thus we do a multipole expansion of the
fields and look at the dipole term.
Then we focus on a patch of the sphere
small enough so that it can be approximated as flat
and
examine the fluctuations in complete generality. We
find the results
of these two calculations reassuring enough that we do
not
compute the stability under higher multipole
perturbations of the
sphere.

For the purposes of this calculation, we write out the
Lagrangian in gory detail and eliminate all hats and
tildes:
\bea
L = \int d\theta d\phi \sin{\theta} \{ \12
(\dot{A_\theta}^2 + \dot{A_\phi}^2)
 -\12 c \left[ {1 \over \sin{\theta}}
(\partial_{\theta} (A_\phi
\sin{\theta}) - \partial_{\phi} A_\theta) \right]^2
+ \12 c^{-1} (\dot D_\theta^2 + \dot D_\phi^2)
\nonumber\\
+ 12\sqrt{2}c^{-\12} \dot{D_\theta} D_\phi
+\12 c^{-1} \dot{\chi}^2
-\12 \left[ (\partial_\theta \chi)^2 +
({{\partial_\phi \chi} \over
\sin{\theta}} )^2 \right] -4\chi^2 \nonumber\\
+ 3\sqrt{6} \epsilon^{ij} A_i \dot{D_j}
- 3\sqrt{6} \chi {1 \over \sin{\theta}} \left[
\partial_\theta(\sin{\theta}
D_\phi) - \partial_\phi D_\theta \right]
+4\sqrt{2} c^{\12} \chi {1 \over \sin{\theta}} \left[
\partial_\theta(\sin{\theta} A_\phi) - \partial_\phi
A_\theta \right]
\}
\eea
where
\be
c = {2 \over 9(\pi N)^{2\over 3} l_s^2}.
\ee
For later convenience, we
have set $\tilde{A_\phi}=A_\phi \sin{\theta}$ and
similarly for
$D_\phi$.

Since the system has rotational symmetry, there is no
harm
in taking the dipole to lie along the $z$ axis.  Then
the
perturbation has azimuthal symmetry so we drop $\phi$
derivatives in the equations of motion:
\bea
-\ddot{A_\theta} + 3 \sqrt{6} c^{-1} \dot{D_\phi}
= 0 \nonumber\\
c \partial_\theta({\partial_\theta(A_\phi
\sin{\theta}) \over
\sin{\theta}}) -
\ddot{A_\phi}
- 4\sqrt{2}c^\12 \partial_\theta \chi -3\sqrt{6}
\dot{D_\theta} =0 \nonumber\\
-c^{-1} \ddot{D_\theta} - 12\sqrt{2} c^{-{1
\over 2}} \dot{D_\phi} + 3 \sqrt{6} {\dot{A_\phi}
\over \sin{\theta}} = 0\nonumber\\
3\sqrt{6} \partial_\theta \chi - c^{-1} \ddot{D_\phi}
+
12 \sqrt{2} c^\12 \dot {D_\theta}
- 3\sqrt{6} {A_\theta}=0\nonumber \\
-c^{-1} \ddot{\chi} + {\partial_\theta(\sin{\theta}
\partial_\theta
\chi)
\over \sin{\theta}} - 8 \chi
 + 4 \sqrt{2} c^\12 {\partial_\theta(\sin{\theta}
A_\phi) \over
\sin{\theta}} - 3\sqrt{6} {\partial_\theta
(\sin{\theta} D_\phi) \over
\sin{\theta}} =
0\nonumber\\
\eea

For the scalar and vector dipole perturbations we are
investigating,
the solutions take the form
\be
A_i (x^\mu) = A{}_i^o e^{i\omega t} \sin{\theta}; \,\
D_i (x^\mu) = D{}_i^o e^{i\omega t} \sin{\theta}; \,\
\chi (x^\mu) = \chi^o e^{i\omega t} \cos{\theta};
\ee
where $A{}_i^o$ is a constant and $i$ runs over
$\theta$ and $\phi$. These are the most general
azimuthally symmetric
scalar and vector dipole harmonics on the sphere.
Plugging in this
ansatz gives:
\bea
 \omega^2 A{}_\theta^o + 3\sqrt{6} c^{-1}
i\omega D{}_\phi^o   =
0\nonumber\\
 ( \omega^2 - 2c )A{}_\phi^o -3\sqrt{6} i
\omega D{}_\theta^o
+ 4\sqrt{2}c^\12\chi^o
= 0
\nonumber\\
 3\sqrt{6} i\omega A{}_\phi^o + c^{-1}
\omega^2D{}_\theta^o -
12\sqrt{2}c^{-{1 \over 2}} i\omega D{}_\phi^o
  = 0
\nonumber\\
 -3\sqrt{6} i\omega A{}_\theta^o +
12\sqrt{2}c^{-{1 \over 2}} D{}_\theta^o +
c^{-1} \omega^2 D{}_\phi^o
- 3\sqrt{6} \chi^o  =0
\nonumber\\
 8\sqrt{2} c^\12 A{}_\phi^o - 6\sqrt{6}
D{}_\phi^o
+ (c^{-1}\omega^2 - 10)\chi^o =0 \nonumber\\
\eea
The resulting system of equations can be put as usual
in matrix form.
In order for there to be nonzero solutions to these
equations, the determinant of the resulting $5 \times
5$ matrix must be
zero. Using this equation to solve for $\omega$ gives
the following
solutions
\be
\omega ^2 ={1 \over{ N^{2 \over 3} l_s^2}}\{ 0, 0,
40.32, 1.37, 0.58 \}
\ee
The nonzero modes have
positive energy and so the system is stable under such
perturbations. There
is no tachyonic mode. The energy scale of the
resulting oscillations
is again of the order of the characteristic energy
scale identified in
the previous sections. The $\omega = 0$ modes are a
bit more complicated.  We will discuss them in a
moment since their complications have nothing to do
with being on a sphere.

Next we examine perturbations with wavelengths smaller
than the size
of the sphere. Rather than doing a higher multipole
expansion, we focus on a small patch of the sphere and
approximate it as flat.
The Lagrangian becomes
\bea
L = \int d^2 z [\12 \dot{\vec{A}}^2 - \12
c (\nabla \times \vec{A})^2 + \12 c^{-1}
\dot{\vec{D}}^2
+ 12\sqrt{2} c^{-\12} \dot{D_1} D_2 + \12
c^{-1} \dot{\chi}^2 \nonumber\\
-\12 (\nabla \chi)^2 - 4 \chi^2
+ 3 \sqrt{6}\epsilon^{ij}A_i \dot D_j  - 3
\sqrt{6} \chi (\nabla \times \vec{D})
+ 4 \sqrt{2} c^\12 \chi (\nabla \times \vec{A})]
\eea
Here the coordinates $z_i$ are dimensionless.
The dimensions
can be restored by multiplying by the radius of the
sphere $R = {(\pi N)^{2/3} l_s/ 2}$. Therefore
the analysis is valid for dimensionless momenta
$p \gg 1$.
We guess a plane wave solution, $A_i =
A{}_i^o e^{i\omega t + i p z_1} $, with similar
expressions for the other fields.  Since $\omega$
will be a function only of  $p^2$, we lose nothing
by assuming the fields are independent of $z_2$.
Then the equations of motion become
\be
\left( \matrix{
        \omega^2 & 0 & 0 & 3\sqrt{6} i \omega & 0 \cr
        0 & \omega^2 - c p^2 & -3\sqrt{6} i \omega & 0
&
-4\sqrt{2} c^\12 i p \cr
        0 & 3 \sqrt{6} i \omega & c^{-1} \omega^2 &
-12\sqrt{2} c^{-\12} i\omega & 0 \cr
        -3\sqrt{6} i \omega & 0 & 12\sqrt{2} c^{-\12}
i
\omega & c^{-1} \omega^2 & 3 \sqrt{6} i p \cr
        0 & 4\sqrt{2} c^\12 i p & 0 & -3 \sqrt{6} i p
&
c^{-1} \omega^2 - p^2 - 8 \cr
        }
\right)
\left( \matrix{
        A_1 \cr
        A_2 \cr
        D_1 \cr
        D_2 \cr
        \chi \cr
} \right) = 0
\ee
The dispersion relations for the normal modes are
shown in figures 1 and 2.

\begin{figure}
\includegraphics[bb=91 560 339
727,clip,scale=1]{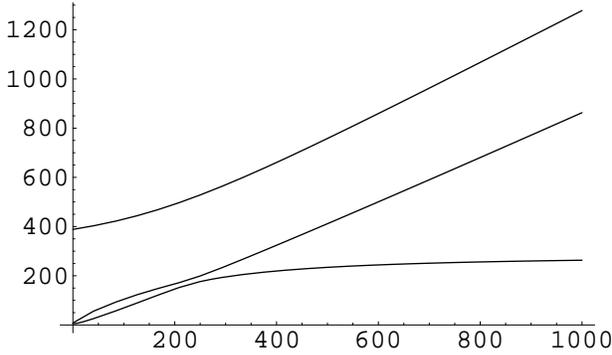}
\caption{The three normal modes of the system;
here we plot $c^{-1} \omega ^2$ as a function of
$p^2$}
\end{figure}

\begin{figure}
\includegraphics[bb=91 560 339
727,clip,scale=1]{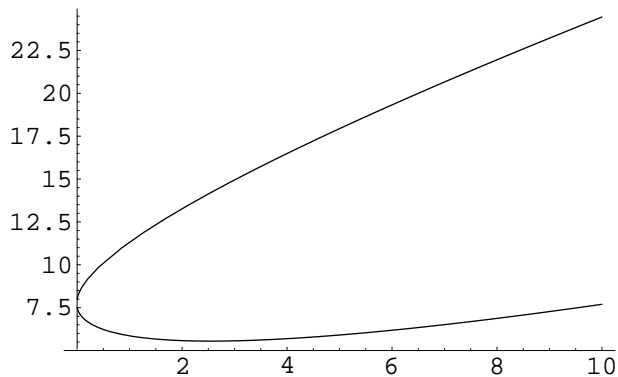}
\caption{A close-up of the two lower modes near $p =
0$;
again we plot $c^{-1} \omega ^2 (p^2)$}
\end{figure}

Since the system is quadratic, when written in terms
of the normal modes it is
a set of uncoupled oscillators whose energies are
quantized as $\omega (n + \12)$ .  However,
the values for $\omega$ given above are not the whole
story because although $\omega = 0$ appears
four times as a solution there are only two
corresponding eigenvectors.  These are pure gauge
and are physically equivalent to $A_i = D_i = \chi =
0$.  The fact that there is not a complete set of
eigenvectors means that we must also consider
solutions which are linear in time\footnote
{Thanks to Boaz Nash for pointing this out and
providing the example which follows.}.
A simple example where such solutions are necessary is
solving
$\ddot{x} = 0$.  If we guess a solution of the form $x
= x_o e^{i \omega t}$ as we did above, then
we find $\omega = 0$ is a double root of the
characteristic equation.  But there is only one
corresponding
eigenvector since it is a one-dimensional space, so
we must supplement our solution with a solution
which is linear in time. Getting back to our problem,
one can picture the solutions this way:
 at each momentum there is a five-dimensional
configuration space; a point in this space represents
a state of the system.
 There is a potential on the space which is a harmonic
oscillator potential in three
of the directions and flat in the other two
directions.  The only subtlety is that the
displacement of the system
along the flat directions can be gauged away; all that
matters is the velocity in these directions.  Also,
the
constraint equation coming from the $A_0$ equation of
motion prevents the system from moving along one of
the flat
directions.  Thus there is only a one-parameter family
of physically distinct non-oscillatory solutions.

{\bf Quasiparticles.}
We will now analyze the non-oscillatory solutions.
These  solutions are linear in time for the potentials
but all physical quantities are constant in time.
For this reason, we find it more convenient to work in
terms of the physical fields.  The genuine electric and
magnetic fields
we denote $E_A$ and $B_A$, while the analogous
quantities for the electron fluid we call $E_D$ and
$B_D$.
These quantities have straightforward physical
interpretations: $B_D$ is proportional to the excess
density
of electrons, while $\epsilon_{ij}E{}_D^j$ is
proportional to the velocity of the electrons.

Writing the equations of motion in terms of the fields
and dropping time derivatives of the fields (since
we're looking
for static solutions) gives the following momentum
space equations:
\bea
i \vec{p} \cdot \vec{E_A} + 3 \sqrt{6} B_D = 0
\nonumber\\
-i c^\12 \vec{p} B_A + 4 \sqrt{2} i \vec{p} \chi - 3
\sqrt{6} c^{- \12} \vec{E_D} = 0 \nonumber\\
3 \sqrt{6} i \vec{p} \chi - 12 \sqrt{2} c^{-\12}
\vec{E_D} + 3 \sqrt{6} \vec{E_A} = 0 \nonumber\\
-(\vec{p}^2 + 8) \chi + 4 \sqrt{2} c^(\12) B_A - 3
\sqrt{6} B_D = 0
\eea
The first of these equations is the constraint from the $A_0$ equation
of motion, eq. (2.85).

As in \cite{susskind}, we are interested in static solutions with
the properties of Laughlin's quasiparticles. In our picture, 
quasiparticles can
be thought of as vortices frozen in the electron fluid. The conserved
quantity (2.88) is therefore non-zero but equal to a time-independent
function appropriate for a localized vortex-type solution.
The static solutions will be quantized. In
terms of the fields, the conserved quantity (in momentum space) is
proportional to
\be
\widetilde{CQ}(\vec{p}) = -12 \sqrt{2} B_D + 3 \sqrt{6} c^\12 B_A + i
c^{- \12} \vec{p} \cdot \vec{E_D}.
\ee
Note that the value of this quantity at each point in
space and not just its integral is conserved.

To get a sensible quantization
condition, we follow \cite{susskind}. From the Lagrangian (2.71), the
canonical momentum density conjugate to $X_s^{i}$ is
\be
\Pi_i = \delta L/ \delta \dot X{}_s^i = -\eta_0 m
g^{00}g_{ij}\partial_t X{}_s^j -
{e \eta_0 B \over 2}\epsilon_{ij}X{}_s^j + e \eta_0
\hat A_i
\ee
Then the momentum per particle $P_i$ is the momentum
density $\Pi_i$ divided by the density of electrons
$\eta_0$.
In order to quantize, we impose the condition
\be
\oint P_i dX^i = 2 \pi (k + \12)
\ee
along any closed path in space.
In terms of the canonically normalized D field, the
positions of the electrons are given by
\be
X{}_s^i = z^i + {12 \sqrt{2} \pi ^ {2/3} l_s^{1/2} \over
{N^{1/3} K^{1/2}}} \epsilon^{ij} D_j.
\ee

Then, to first order, the quantization condition becomes
\be
B \times area + \int dz_1 dz_2  {\pi ^ {2/3} N^{2/3} l_s^{1/2} \over
{2 K^{1/2}}} CQ(z) = 2 \pi (k + \12)
\ee
where $k$ is an integer, and we have turned the line
integral into a volume integral.  The zeroth order
term gives the quantization of the magnetic flux; the first order
term gives the quantization of the quasiparticle
excitations.

The quantization condition for the quasiparticles thus
becomes
\be
\int dz_1 dz_2 CQ(z) = { 4 \pi^{1/3} K^{1/2} \over {
N^{2/3} l_s^{1/2}}} k
\ee
As in \cite{susskind}, we interpret this equation to
mean that the introduction of quasiparticles can only
change $k$ by an integer, regardless of the region of
integration.  Thus we choose for our quasiparticle a
delta function of quantized strength. This will lead in general to
to field configurations that are singular at the position of the
vortex. As noted in
\cite{susskind}, the divergence is smoothed out if
terms higher order in the fields are included, leading to non-linear
terms in the equations; we do
not find the divergence annoying enough to smooth it
out.
The smallest size quasiparticle
sitting at the origin is given by $CQ(z) =  ({ 4
\pi^{1/3} K^{1/2}/ { N^{2/3} l_s^{1/2}}})
\delta^2(z) $.
In momentum space, this condition becomes
\be
\widetilde{CQ}(\vec{p}) = -12 \sqrt{2} B_D + 3
\sqrt{6}
c^\12 B_A + i c^{- \12} \vec{p} \cdot \vec{E_D} =
 { 4 \pi^{1/3} K^{1/2} \over { N^{2/3} l_s^{1/2}}}
\ee

Solving the equations of motion in
momentum space, we find the solution
\bea
\vec{E_A} = { 4 \pi^{1/3} K^{1/2} \over { N^{2/3}
l_s^{1/2}}} {i \sqrt{3} (6 - p^2) \vec {p} \over {2 (162
+ 15 p^2 + 2 p^4)}} \nonumber\\
\vec{E_D} = { 4 \pi^{1/3} K^{1/2} c^{\12} \over {
N^{2/3}
l_s^{1/2}}}   {i 9 \vec {p} \over {162 + 15 p^2 +
2 p^4}} \nonumber\\
B_A = { 4 \pi^{1/3} K^{1/2} c^{-\12} \over { N^{2/3}
l_s^{1/2}}}
 {(9 + 2 p^2) \sqrt{6} \over {162 + 15 p^2 + 2
p^4}} \nonumber\\
B_D=  { 4 \pi^{1/3} K^{1/2} \over { N^{2/3} l_s^{1/2}}}
{6 p^2 - p^4  \over {6 \sqrt{2} (162 + 15 p^2 + 2
p^4)}} \nonumber\\
\chi = { 4 \pi^{1/3} K^{1/2} \over { N^{2/3} l_s^{1/2}}}
{(18 + p^2) \sqrt{3} \over {2 (162 + 15 p^2 + 2 p^4)}}
\eea
Note that such a solution is not really allowed on the
sphere because the total number of electrons
is not conserved; our real interest is forming a
particle/hole pair. To understand the above solution,
note that $B_D$ is the most badly behaved at large $p$
and thus at small $z$; the singular part is
proportional to a delta function in position space.  Thus there is
a finite charge at a point in $z$-space.  As discussed
in \cite{susskind}, this point actually takes up a finite
area in real space.  What is going on in the case of a
hole is that the electron fluid is getting moved
radially outward to make the hole.  We find that the
quasiparticle has charge $\nu$ as expected. To see this, note that the
change in the density of electrons in real space is given by
\be
\delta\eta(z) =  {3 \sqrt{2} \pi ^ {-1/3} l_s^{1/2}K^{1/2} \over
{N^{1/3} }} B_D(z) = -\nu \delta^2(z) + ....
\ee
Thus there is a hole of electrons at $z=0$ with charge $\nu$.
The only other field
that blows up is $E_A$, but we expect the electric field to
blow up near a point-like charge.  $E_D$ is radial, which by
(2.3) means that the electrons are moving in the angular
direction. The magnetic force arising due to the motion of the
electrons balances the Coulomb force and the force due to the scalar
field. Thus the quasiparticle is a vortex in the electron
fluid.

We would like to compute the energy of
a single quasiparticle.  The Hamiltonian in
momentum space is
\be
\int {d^2 p \over {(2 \pi)^2}} [ \12 |E_A|^2 + \12 c
|B_A|^2  +  \12 c^{-1} |E_D|^2
+ \12 (p^2 + 4) |\chi|^2  + 3 \sqrt{6} {\chi^* B_D +
\chi B_D^* \over 2} +  4 \sqrt{2} {\chi^* B_A + \chi
B_A^* \over 2} ]
\ee
which for this solution magically simplifies to
\be
\int {d^2 p \over {(2 \pi)^2}} {16 K  \pi^{2/3} \over
{N^{4/3}l_s}} {-9 \over {2 (162 + 15 p^2 + 2 p^4)}}
\ee
Note that this is clearly negative, and also that we
don't need to impose an ultraviolet cutoff; although
various fields blow up at large $p$, they cancel in
such a way that the energy is finite.  Physically, this
is because the forces between electrons at short
distances become small due to cancellations between
the scalar and the electromagnetic field
\cite{gubser}.

The result of integrating gives
\be
E_{qp} = -{1.26 K \over \pi^{1/3} N^{4/3} l_s} =
-{1.26 \nu \over (\pi N)^{1/3} l_s}
\ee
This energy is of the scale predicted in
\cite{giantbob}
but it is negative!

Because the equations of motion are linear, if we can
have one quasiparticle we can have as many as we want
at any locations we want. The quasiparticles do not
move in this solution.
In order to have a quasiparticle/quasihole pair, one
at the origin and one at $x_0$, we multiply the right
side of the equation for the conserved quantity in
momentum
space by $(1 - e^{i \vec{p} \cdot \vec{x}_0})$.
Clearly
a particle-hole pair at zero separation has zero
energy,
and a pair at large separation has twice the energy
of one quasiparticle, so the energy is lowered by
creating
particle-hole pairs and separating them. However, we note again that
there are static solutions with the particle and hole at any 
separation.


This seems like an instability.  However, note that
these are all viable static solutions and have
different values of the conserved quantitity.  The
issue is whether
adding higher-order corrections to the Lagrangian will
 allow the system to move to a
state with more quasiparticles, lowering its energy.
We speculate about whether this happens in the
conclusions.

\setcounter{equation}0
\section{Conclusions}

We have shown unambiguously that the Quantum Hall
Soliton
is stable to small perturbations and that it contains
fractionally charged excitations.  The puzzling
question
is whether the negative energy associated with a
quasiparticle
indicates an instability.

One possibility is that the symmetry which leads to
conservation
of quasiparticles will remain good when we add higher
terms to
the Lagrangian.
The gauge symmetry which leads to the conserved
quantity is the symmetry of the electron fluid under
area-preserving diffeomorphisms.  When discretized,
this symmetry becomes a $U(K)$ symmetry.  It was
conjectured in \cite{susskind} that electrons in a
magnetic field have an exact $U(K)$ symmetry when the
kinetic term is dropped.  We suspect that here because
the electrons are connected to strings, they only
truly have the permutation symmetry and not the
full $U(K)$ symmetry.
What could happen for example is that the strings
could become excited while the brane goes into a
lower- energy configuration with more quasiparticles.

We do not believe that this is actually an instability
for the following reason.
Let us return to our example of a particle subject to
an upside-down
oscillator potential and a magnetic field
and continue to ignore the kinetic term.
A solution exists for the particle orbiting at any
radius, and
the larger the radius the more negative the energy.
Now consider
weakly coupling the particle to an oscillator with
higher
frequency by the coupling $\epsilon \vec{x}_{part}
\cdot \vec{x}_{osc}$.
This coupling destroys angular momentum conservation
and one might think it would allow the system to slide
down
to lower and lower energy states, giving energy to the
oscillator.
But the only effect of the coupling is to slightly mix
the
oscillator mode with the cyclotron mode, INCREASING
the
frequency of the cyclotron mode.  We believe that this
example
is strongly analagous to the situation of a stable
brane with
negative energy states coupled to some strings, and
for this
reason we believe the effect of adding string
oscillations will
be to slightly change the frequencies of the normal
modes we have found.
In particular, we expect it to give our static
solutions a small
positive $\omega^2$.

We briefly mention other possible ways the $D2$-brane
could give up
its energy.  The $D6$-brane is infinite in $6$
directions and thus has
a large number of low-frequency modes. These
basically act as a
frictional force on the electrons and will eventually
steal the energy.
As was discussed in \cite{giantbob}, we could
compactify the directions
parallel to the $D6$-brane to remove these modes.
Finally, the $D2$-brane could emit closed string
gravitons.  Again, in the large $N$ limit we are
working in, the ``closed string'' coupling constant is small and the 
time scale
for this process is large.

Unfortunately, the limitations of our method prevent
us from making these arguments more rigorous.

\setcounter{equation}0
\section{Appendix}

In this Appendix, we review the relation between the
``Eulerian'' and
``Lagrangian'' descriptions of the $D0$-brane fluid
following
\cite{cornalba}\cite{seiberg}.

Let us begin with the ``Eulerian'' description. We
choose to work in
the $A_0=0$ gauge and split the gauge field into its
background value
and the fluctuations
\be
A_i(X)=-{1 \over 2}B_{ij}X^j + A^f_i.
\ee
The field strength is given by
\be
F_{0i}=\partial_tA_i=\partial_tA^f_i
\ee
and
\be
F_{ij}=B_{ij}+F^f_{ij}.
\ee
Next we do a coordinate transformation of the form
\be
X^0=t, \,\  X^i=y^i + \Theta^{ij}\hat{A_j}(y,t)
\ee
so that
\be
\tilde{F}_{0i}=F_{0j}{\partial X^j \over \partial
y^i}+F_{kl}{\partial
X^k \over \partial t}{\partial X^l \over \partial
y^i}= 0
\ee
and
\be
\tilde{F}_{ij}=F_{kl}{\partial
X^k \over \partial y^i}{\partial X^l \over \partial
y^j}= B_{ij}.
\ee
Eq. (6.6) can be satisfied to quadratic order in the
fluctuations if
we set
\be
\hat{A_i}=A^f_i + \12
\Theta^{kl}(2A^f_l\partial_kA^f_i +
A^f_k\partial_iA^f_l)+ h.o.
\ee
This is the Seiberg-Witten map between ordinary and
non-commutative
gauge fields to leading order in $\Theta$.
In this way the coordinate frame $t, \,\ y^1, \,\,
y^2$ is determined
up to infinitesimal area preserving diffeomorphisms of
the form
\be
y'^i=y^i + \Theta^{ij}\partial_j \hat{\lambda}(t, y).
\ee
Under such a tranformation the non-commutative gauge
field transforms
as in eq. (2.15) but $\tilde{F}_{ij}=B_{ij}$ remains
unchanged. To
satisfy equation (6.5) then, we do a particular
time-dependent area
preserving diffeomorphism. To satisfy
$\tilde{F}_{0i}=0$ to quadratic
order in the fluctuations, we must set
\be
\hat{A_i} \rightarrow \hat{A_i} +
\partial_i\hat{\lambda}
\ee
with $\hat{\lambda}$ satisfying
\be
\partial_t \hat{\lambda} = \12 \Theta^{ij}
A^f_i\partial_tA^f_j+ h.o.
\ee
Still, there is left over coordinate freedom
consisting of time
independent area preserving diffeomorphisms.

Alternatively, we may leave $A_0$ to be non-zero in
the Eulerian
description. Then we can satisfy eq. (6.5) for
$\hat{A_i}$ given by
eq. (6.7) if we set
\be
A_0 = -\12 \Theta^{ij} A^f_i\partial_tA^f_j+h.o.
\ee
By doing then a time dependent commutative gauge
transformation, we
can reach again the $A_0=0$ gauge. Thus we must set
\be
\partial_t{\lambda} = \12 \Theta^{ij}
A^f_i\partial_tA^f_j +h.o.
\ee
The two gauge parameters are related to each other as
in the
Seiberg-Witten map
\be
\hat{\lambda}=\lambda + O(\Theta).
\ee

\section{Acknowledgements}
We would like to thank Boaz Nash and John McGreevy
for helpful discussions. This work was supported in part by NSF grant 
980115.
B.F. is supported in part by an NSF Graduate
Fellowship.

\end{document}